# Bulk superconductivity in single phase $Bi_3O_2S_3$

**Jifeng Shao[1], Zhongheng Liu[1], Xiong Yao[1], Li Pi[1], Shun Tan[1], Changjin Zhang[*,1] and Yuheng Zhang[1]**

[1] High Magnetic Field Laboratory, Chinese Academy of Sciences and University of Science and Technology of China, Hefei 230026, People's Republic of China

We report the synthesis of single phase $Bi_3O_2S_3$ sample and confirm the occurrence of bulk superconductivity with transition temperature at 5.8 K. The $Bi_3O_2S_3$ superconductor is categorized as the typical type-II superconductor based on the results of both temperature and magnetic field dependences of magnetization. Hall coefficient measurements give evidence of a multiband character, with a dominant conduction mainly by electron-like charge carriers. The charge carrier density is about $1.45 \times 10^{19}$ cm$^3$, suggesting that the system has very low charge carrier density.

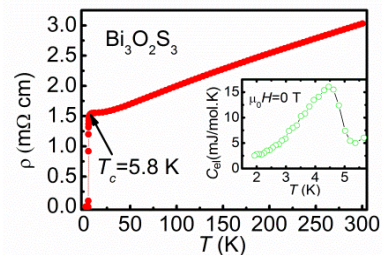

This is the caption of the **optional** abstract figure. If there is no figure here, the abstract text should be divided into both columns.

**1 Introduction** Materials with layered structure have been intensively studied as a promising approach to the exploration of new superconductors, since the discovery of cuprate superconductors [1]. This approach has become more convincing due to the discovery of iron-based superconductors [2], which have a layered structure consisting of the so-called superconducting layers ($Fe_2M_2$ layers, M=P, As, Se, or Te) and blocking charge reservoir layers [3, 4]. Superconductivity occurs when charge carriers are generated in blocking layers and transferred into the superconducting layers.

Recently, $BiS_2$-based superconductors have attracted considerable attention [5-16]. The first $BiS_2$-based superconductor is $Bi_4O_4S_3$, which is composed of $Bi_2O_2(SO_4)_{0.5}$ blocking layers and $BiS_2$ superconducting layers [5]. Subsequently, a series of $BiS_2$-based superconductors have been found, such as $ReO_{1-x}F_xBiS_2$ (Re= La, Ce, Nd, Yb, Pr) [5-10], $La_{1-x}M_xOBiS_2$ (M=Ti, Zr, Hf, Th) [11], $Sr_{1-x}La_xFBiS_2$ [12], and so on.

In a recent study of the Bi-O-S systems, Phelan et al. have found two new Bi-O-S phases, with the chemical formula of $Bi_2OS_2$ and $Bi_3O_2S_3$, respectively [13]. They claim that if the majority phase is $Bi_3O_2S_3$, the sample exhibits superconducting transition at 4.5 K. The superconductivity is rapidly suppressed as the $Bi_2OS_2$-like stacking faults are introduced in $Bi_3O_2S_3$. This report is interesting in the exploration of new superconducting materials. However, due to the lack of single phase sample, the superconductivity in $Bi_3O_2S_3$ remains to be controversial. In this Letter, we report the successful synthesis of single phase $Bi_3O_2S_3$ and bulk superconductivity with transition temperature at 5.8 K.

**2 Experimental** Polycrystalline sample of $Bi_3O_2S_3$ was synthesized by a conventional solid state reaction method. High-purity starting materials of $Bi_2O_3$ and S were weighed with nominal composition $Bi_{12}O_{18}S_{13}$ and thoroughly ground in an agate mortar. It should be noted that the nominal S content is higher than that in ref. 13. The mixed power was pressed into pellets, sealed in evacuated quartz tubes and then placed in a preheated furnace at 430°C. After being heated for 2 h, the sample was quenched down to 0°C by putting the quartz tubes into ice water. It should be very careful when the products were removed from the sealed quartz tubes as large amount of S-O gas (such as $SO_2$ and $O_2$) was produced during heating.

The products were mixed by regrinding, which were pelletized again and sealed in evacuated quartz tubes. An intermediate preheat was performed at 430°C for 15 h and then quenched down to 0°C. After another regrinding and pelletizing, the sample was sealed in evacuated quartz tubes and sintered at 520°C for 17 hours. The obtained sample is hard and black, stable in air.

The real composition of the obtained sample is determined using an energy dispersive x-ray spectroscopy (EDX) analysis, which was performed using Oxford SWIFT3000 spectroscopy equipped with a Si detector. The lattice structure of the sample was characterized by powder x-ray diffraction with Cu-K$_\alpha$ at room temperature. The temperature dependence of resistivity from 2K to 300K was measured by a standard four-probe method in a commercial Quantum Design PPMS-14T system. Magnetic properties were performed using a superconducting quantum interference device (SQUID) magnetometer.

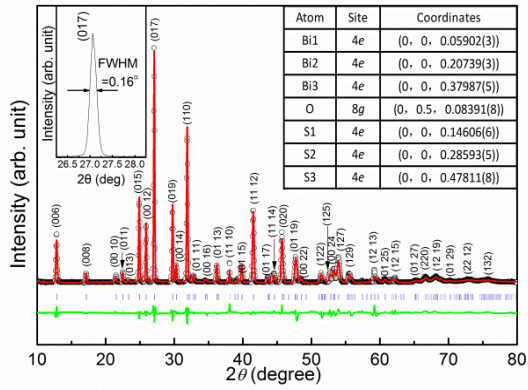

**Figure 1** Powder x-ray diffraction pattern of the $Bi_3O_2S_3$ sample (black dots) and its Rietveld refinement (red curve). The inset shows the obtained site occupancy.

**3 Results and discussion** Figure 1 shows the powder x-ray diffraction (XRD) pattern of the as-grown sample as well as the Rietveld refinement ($R_{WP}$ =8.4 %). It can be seen from Fig. 1 that all the diffraction peaks can be indexed according to the space group of $I4/mmm$. Rietveld refinement on the XRD pattern gives the lattice constants of $a$ = 3.9674 Å and $c$ = 41.2825 Å. The obtained atomic parameters and the site occupancy are summarized in the inset of Fig. 1. An enlarged view of the (017) diffraction peak gives no evidence of broadening (the full width at half maximum is 0.16°). And there is no diffraction peak at around 37.6°. These facts exclude the possibility of the presence of Bi impurity. We also notice that other possible impurities, such as $Bi_2O_3$ and $Bi_2S_3$, do not give substantial contribution to the XRD pattern. It should be mentioned that the lattice constants of the $Bi_3O_2S_3$ phase is similar to those of $Bi_4O_4S_3$ [5]. We make a refinement of the XRD data using the lattice constants and atomic coordinates of $Bi_4O_4S_3$ taken from Ref. 5. In the refinement, the lattice constants and atomic coordinates are adjusted one by one. The obtained $R_p$(23.1%) and $R_{wp}$(30.22%) value exceed the confidence interval, indicating that the sample is not of $Bi_4O_4S_3$ phase.

We make a careful analysis on the chemical composition of the sample using an EDX analysis. The sintered specimen is manually broken into many small pieces. We choose twelve pieces in EDX analysis and the average is determined as the real composition. Since our Oxford SWIFT3000 spectroscopy cannot give quantitative information on oxygen content, we analyze only the relative proportion of Bi and S. It is found that for all the chosen pieces, the relative ratio of Bi and S is close to 1:1. The EDX analysis results confirm that the chemical composition of the sample is $Bi_3O_2S_3$.

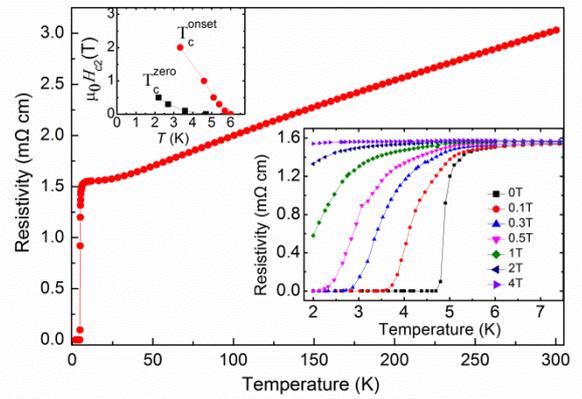

**Figure 2** Temperature dependence of resistivity for the $Bi_3O_2S_3$ sample. The lower inset shows the resistivity for the $Bi_3O_2S_3$ sample under different applied magnetic field. The upper inset shows the field dependence of $T_c^{onset}$ and $T_c^{zero}$.

The temperature dependence of resistivity of the $Bi_3O_2S_3$ sample is given in Fig. 2. It exhibits metallic-like behavior below 300 K, which is similar to that of $Bi_4O_4S_3$ [5]. A sharp drop appears below 5.8 K, which is corresponding to the superconducting transition. In order to investigate the superconducting properties under magnetic field, we perform the resistivity versus temperature measurement with applied magnetic field. The results are shown in the inset of Fig. 2. In can be seen that the onset of the superconducting transition temperature decreases and the width of transition becomes large with increasing magnetic field, meaning that the superconducting states are destroyed by applying magnetic field. The variation of $T_c^{onset}$ and $T_c^{zero}$ values with applied magnetic fields is shown in the upper inset of Fig. 2. The upper critical field $\mu_0H_{c2}(0)$, determined using the Werthamer-Helfand-Hohenberg (WHH) formula $H_{c2}$ = -0.69$T_c$ [$dH_{c2}/dT$ ]$_{Tc}$ [17], is about 4.84 T.

The temperature dependence of magnetic susceptibility is shown in Fig. 3. The measurement is performed on the ground powder specimens to avoid suffering from shield-ing of empty voids in the pellets. A large diamagnetic signal is observed below 5.1 K, confirming the occurrence of superconducting transition. We calculate the shielding volume fraction (SVF) at 2K by SVF=$4\pi M\rho/H$, where $M$ is the magnetization at emu/g scale, $\rho$ is the density at the unit of g/cm$^3$, and $H$ is the applied magnetic field at the unit of Oe. The SVF value is 96.5%. The divergence of temperature dependence of magnetic susceptibility measured under magnetic-field-cooling process and under zero-field-cooling process gives the indication of type-II superconductor for $Bi_3O_2S_3$.

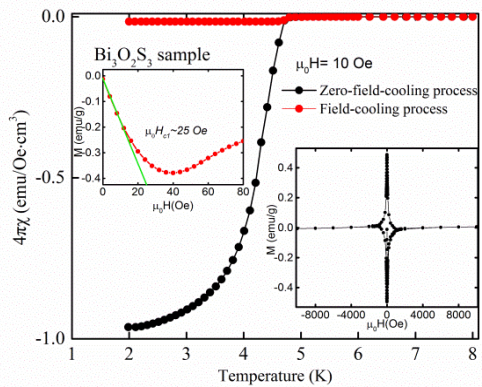

**Figure 3** Temperature dependence of magnetic susceptibility for the $Bi_3O_2S_3$ sample. The right inset shows the magnetic field dependence of magnetic susceptibility at 2K. The left inset shows the low-field region data of the $M\sim H$ curve.

Figure 4 gives the results of specific heat ($C_P$) in the superconducting state ($\mu_0H=0$ T) and normal-conducting state ($\mu_0H=9$ T). The jump associated with superconducting transition is observed, confirming the occurrence of bulk superconductivity in $Bi_3O_2S_3$. The tiny jump is attributed to the small electronic-specific-heat coefficient $\gamma$. The specific heat can be expressed as $C_P(T) = \gamma T + \beta T^3$, where $\gamma$ is the electron contribution to the specific heat and $\beta$ is the phonon contribution. The obtained values of $\gamma$ and $\beta$ are 1.65 mJ mol$^{-1}$ K$^{-2}$ and 2.6 mJ mol$^{-1}$ K$^{-4}$, respectively. The Debye temperature is estimated to be 182 K. The phononic contribution to the heat capacity is generally independent on applied magnetic field. Thus we calculate the electronic specific heat under zero field $C_{el}(T)$ by substracting the phonon contribution: $C_{el}(T) = C_P(T, H=0) - C_P(T, H=9\text{ T}) + \gamma T$, where $\gamma$ is the electronic contribution to the specific heat determined from the fitting of low temperature $C_P(T, H=9\text{ T})$ data. The result is shown in the right bottom inset of Fig. 4, from which the jump in electronic specific heat at $T_c$ = 4.9 K can be clearly seen. The value of $\Delta C_e/\gamma T_c$ is 1.42, which is comparable to BCS weak-coupling limit value 1.43.

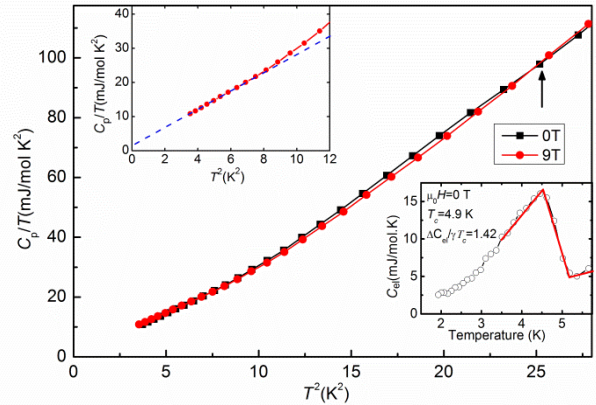

**Figure 4** Temperature dependence of specific heat of the $Bi_3O_2S_3$ sample. The top inset shows the $C_P/T$ vs $T^2$ plot in the low-temperature region at 9 T (normal state). The inset at right bottom shows the calculated $C_{el}\sim T$ curve under zero field.

The magnetic field dependence of magnetization ($M\sim H$ curve) measured at 2K for the $Bi_3O_2S_3$ sample is given in the inset of Fig. 3. The lower critical field ($\mu_0H_{c1}$), defined as the field at which the $M\sim H$ curve deviates from the initial linear slope, is 25 Oe at 2 K. According to the Ginzburg-Landau (GL) formula $H_c(T)=H_c(0)[1-(T/T_c)^2]$, the lower critical field at 0K is about 30.5 Oe. As estimated above, the upper critical field at 0 K is $H_{c2}(0)$=48400 Oe. Thus the GL parameter $\kappa(0)$ is estimated to be 56.5 using the equation $H_{c2}(0)/H_{c1}(0) = 2\kappa(0)^2/\ln\kappa(0)$. The GL coherence length at 0 K ($\xi(0)$) is about 8.24 nm, estimated by $\xi(0)=\{\Phi_0/[2\pi H_{c2}(0)]\}^{1/2}$. And the GL penetration depth $\lambda(0)$ is 466 nm according to the equation $\kappa(0)= \lambda(0)/\xi(0)$.

In order to determine the type and density of charge carriers, we perform the measurements of Hall coefficient. The transverse resistivity $\rho_{xy}$ gives a negative value at all temperature (Fig. 5(a)), suggesting that electron-type charge carriers give dominant contribution to the charge transport. A strong nonlinearity of the Hall effect has been observed, suggesting that the sample cannot be described by a simple single band description of Fermi surface. Similar nonlinearity of the Hall effect has also been found in epitaxial $MgB_2$ thin films, a well-known multiband superconductor [18]. In $Bi_4O_4S_3$ superconductor, the nonlinearity of the Hall effect has lead to the argument of multiband character [19], which is further supported by a muon-spin spectroscopy study [20]. Due to the nonlinearity of the Hall effect, it is not possible to determine the Hall coefficient ($R_H$) using the slope of $d\rho_{xy}/dH$, we thus determine the $R_H$ by using the formula $R_H = \rho_{xy}/H$ at fixed magnetic field. The temperature dependence of $R_H$ at different magnetic field is shown in Fig. 5(c). We notice that the normal state $\rho_{xy}\sim\mu_0H$ curves of the $Bi_3O_2S_3$ sample exhibit quasi-linear behavior at low magnetic field (e.g. <1 T), as shown in Fig. 5(b). Thus it can be concluded that the $R_H$ value is independent of magnetic field when $\mu_0H<1$ T. Due to this fact, we estimate the charge carrier density of $Bi_3O_2S_3$ sample at low magnetic field using $n=1/R_He$. The variation of charge carrier density with temperature is plotted in Fig. 5(d). The charge carrier density is close to $1.45 \times 10^{19}$ cm$^{-3}$, which is much lower than those of cuprate superconductors and Fe-based superconductors, implying that the superfluid density in $Bi_3O_2S_3$ superconductor is diluted.

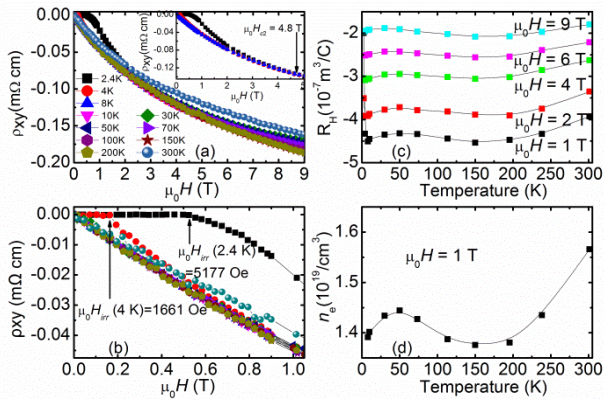

**Figure 5** (a) Transverse resistivity $\rho_{xy}$ versus magnetic field $\mu_0H$ at different temperatures. (b) An enlarged view of the $\rho_{xy} \sim \mu_0H$ curves at $\mu_0H<1$ T. (c) The Hall coefficient $R_H$ at different magnetic field. (d) Estimated charge carrier concentration.

Layered materials have become the most promising candidate in exploration of new superconductors. For examples, the cuprate superconductors and iron-based superconductors are all composed of superconducting layers and various charge reservoir layers. The cuprate superconductors are mainly composed of five different superconducting families: the $La_{2-x}M_xCuO_4$ (M=Ba, Sr, etc.) family, the $RE_{2-x}Ce_xCuO_4$ (RE=Nd, Pr, Sm, Gd, Eu) family, the $REBa_2Cu_3O_{7-\delta}$ family, the Bi-Sr-Ca-Cu-O family, and the Tl(Hg)-Ba-Ca-Cu-O family [21]. Similarly, the iron-based superconductors are mainly composed of five superconducting families: the FeSe(Te) family, the Na(Li)$_{1-x}$FeAs family, the Ba(Sr)Fe$_2$As$_2$-based family, the REFeAsO$_{1-x}$F$_x$ and REFeAsO$_{1-y}$ (RE=La, Sm, Pr, Nd, Ce) family, and the intercalated Fe$_2$Se$_2$ family [3]. For the BiS$_2$-based superconductors, the REO$_{1-x}$F$_x$BiS$_2$ (RE=La, Ce, Nd, Pr) family has become a major group of superconducting candidate. Besides, other superconducting compounds share similar lattice structures have been reported, including Sr$_{1-x}$La$_x$FBiS$_2$ [12], LaO$_{1-x}$F$_x$BiSe$_2$ [22], etc. Here we emphasize that there could be another Bi-O-S based superconducting family. At present, the members of this family include Bi$_4$O$_4$S$_3$ [5], Bi$_6$O$_8$S$_5$ [23], and Bi$_3$O$_2$S$_3$. It is hopeful that more superconducting materials based on this Bi-O-S family will be discovered in future.

**Acknowledgements** This work was supported by the State Key Project of Fundamental Research of China (Grant Nos. 2010CB923403 and 2011CBA0011) and the Natural Science Foundation of China (Grant Nos. 11174290 and U1232142).